\documentclass[%
 reprint,
superscriptaddress,
nofootinbib,
 amsmath,amssymb,
 aps,
 prl,
floatfix,
]{revtex4-2}

\usepackage{enumitem}

\usepackage{graphicx}
\usepackage{dcolumn}
\usepackage{bm}

\usepackage{epsfig}
\usepackage{amsmath}
\input{epsf}
\usepackage{psfrag}
\usepackage{xcolor}
\usepackage{pdfpages}
\usepackage[normalem]{ulem}
\usepackage{soul}
\makeatletter
\AtBeginDocument{\let\LS@rot\@undefined}
\makeatother
\newcommand{\bea}{\begin{eqnarray}}
\newcommand{\eea}{\end{eqnarray}}

\newcommand{\nn}{\nonumber}

\begin{document}

\title{ Long Range Azimuthal Correlation, Entanglement and Bell Inequality Violation by Spinning Gluons at the LHC}

\author{Yuxun Guo}%
 \email{yuxunguo@lbl.gov}
\affiliation{Nuclear Science Division, Lawrence Berkeley National Laboratory, Berkeley, CA 94720, USA}%

\author{Xiaohui Liu}
 \email{xiliu@bnu.edu.cn}
 \affiliation{Center of Advanced Quantum Studies, School of Physics and Astronomy, Beijing Normal University, Beijing, 100875, China}
 \affiliation{Key Laboratory of Multi-scale Spin Physics, Ministry of Education, Beijing Normal University, Beijing 100875, China}

\author{Feng Yuan}%
 \email{fyuan@lbl.gov}
\affiliation{Nuclear Science Division, Lawrence Berkeley National Laboratory, Berkeley, CA 94720, USA}%
\affiliation{Institute for Theoretical Physics,
                Universit\"{a}t T\"{u}bingen,
                Auf der Morgenstelle 14,
                D-72076 T\"{u}bingen, Germany}

\author{Hua Xing Zhu}%
 \email{zhuhx@pku.edu.cn}
\affiliation{School of Physics, Peking University, Beijing 100871, China}%
\affiliation{Center for High Energy Physics, Peking University, Beijing 100871, China}

\begin{abstract}
We apply the recently developed concept of the nucleon energy-energy correlator (NEEC) for the gluon sector to investigate the long-range azimuthal angular correlations in proton-proton collisions at the LHC. The spinning gluon in these collisions will introduce a significant nonzero $\cos(2\phi)$ asymmetries in both Higgs Boson and top quark pair productions.  
The genesis of the $\cos(2\phi)$ correlation lies in the intricate quantum entanglement. Owing to the substantial $\cos(2\phi)$ effect, the NEEC observable in Higgs Boson and $t{\bar t}$ production emerges as a pivotal avenue for delving into quantum entanglement and scrutinizing the Bell inequality at high-energy colliders.  \end{abstract}

\maketitle

\textbf{\textit{  Introduction.}} Long range correlation in particle productions in proton-proton ($pp$) collisions at the LHC has attracted great attention in the last decade with tremendous efforts from both experiment and theory sides~\cite{Dusling:2015gta,Loizides:2016tew,Strickland:2018exs,Nagle:2018nvi}. In this paper, we investigate this physics from a different perspective, applying the nucleon energy-energy correlator (NEEC)~\cite{Liu:2022wop,Cao:2023oef,Li:2023gkh} at the LHC. We will show that the spinning gluon distribution in this framework~\cite{Li:2023gkh} leads to sizable $\cos(2\phi)$ azimuthal asymmetries in forward-backward energy correlators in $pp$ collisions. 
These long range $\cos(2\phi)$ asymmetries are signatures of the quantum entanglement, thereby providing the first test of the Bell Inequality~\cite{Bell:1964kc,Clauser:1969ny} within the entangled gluon system. Pursuing such a test in the Standard Model (SM) of particle physics at high energy colliders has been very active in recent years~\cite{Fabbrichesi:2021npl,Severi:2021cnj,Barr:2021zcp,Aguilar-Saavedra:2022uye,Aguilar-Saavedra:2022mpg,Ashby-Pickering:2022umy,Fabbrichesi:2023cev,Aguilar-Saavedra:2023hss,Bi:2023uop,Han:2023fci,Ma:2023yvd,Bernal:2024xhm,Barr:2024djo,Maltoni:2024csn}.

The NEEC was introduced in~\cite{Liu:2022wop} as a new method to explore the nucleon structures. 
It employs an asymptotic energy flow operator $\hat{{\cal E}}(\theta_a)$ which measures energy deposits in the detector at a fixed angle $\theta_a$ relative to the nucleon incoming beam direction in collider experiments. Previous studies mainly focused on the deep inelastic scattering (DIS)~\cite{Liu:2022wop,Cao:2023oef,Liu:2023aqb,Li:2023gkh} which will be explored at the future electron-ion collider (EIC)~\cite{Accardi:2012qut,AbdulKhalek:2021gbh,Proceedings:2020eah}. In the following, we will study the NEEC observables in $pp$ collisions. The comparison between these two collision systems will provide an opportunity to test the universality of the NEECs. Meanwhile, the novel phenomena unveiled below will stimulate further experiment investigations and help decipher the origin of nearside ridge in $pp$ collisions. 

\begin{figure}[htbp]
  \begin{center}
   \includegraphics[scale=0.85]{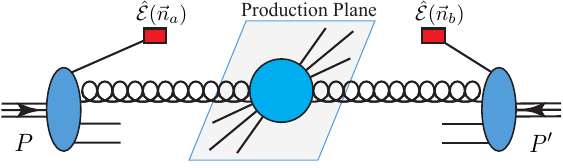} 
\caption{Nucleon energy-energy correlator measurements in proton-proton collisions at the LHC. Energy deposits in the forward directions of both incoming hadron beams with polar angles $\theta_{a,b}$ and azimuthal angles $\phi_{a,b}$ represented by $\vec n_{a,b}$, respectively. 
}
  \label{fg:measure}
 \end{center}
  \vspace{-5.ex}
\end{figure}

To investigate the NEEC at the LHC, we propose to measure the energy deposits along the beam directions of incoming hadrons with polar angles $\theta_{a,b}$ and azimuthal angles $\phi_{a,b}$, respectively, see the illustration in Fig.~\ref{fg:measure}. The hard partonic scattering produces, e.g., the Higgs Boson, top quark pair. The experiment can be carried out by a coincidence measurement between the forward/backward energy flows and the hard interactions in the central. 
Because $\theta_{a}$ and $\theta_b$ are both small and in opposite directions, their rapidity difference will be large, for which we refer to as a long range correlation. Meanwhile, we will show that different processes lead to different $\cos(2\phi)$ asymmetries. Especially, we find that the asymmetries in Higgs Boson and top quark pair productions are quite sizable but with opposite signs. Therefore, a detailed study of these correlations will open a new avenue for precision SM physics. 

In the following, we focus on the gluon NEEC~\cite{Li:2023gkh},
\bea\label{eq:fgmunu}
&& f^{\alpha\beta}_{g,{\rm EEC}} (x,\vec{n}_a)
= 
\int \frac{dy^-}{2\pi xP^+ } e^{- i x P^+ \frac{y^-}{2} }   \nn \\ 
&&
\hspace{4.ex} 
\times  
\langle P| 
{\cal F}^{+\alpha}
\left(y^- \right) 
{\cal L}^\dagger[\bm{\infty},y^-]
\hat{{\cal E}}({\vec n}_a)   
{\cal L} [\bm{\infty},0]
{\cal F}^{+\beta}
(0)  |P \rangle  \,
\nn \\ 
&&
= \left(-{g_T^{\alpha\beta}}/{2}\right) f_{g,{\rm EEC}}
+ h_T^{\alpha\beta}
d_{g,{\rm EEC}}\,,
\eea 
for the proton moving in $+\hat z$ direction with momentum $P$, where ${\cal F}$ is the gauge field strength tensor, and ${\cal L}$ is the gauge link. We have kept the azimuthal dependence of the energy flow direction $n_a^\alpha = (1,\sin\theta_a \cos\phi_a, \sin\theta_a \sin \phi_a, \cos\theta_a)$. To parameterize the spinning gluon distribution, we introduced two projection tensors: $g_T^{\alpha\beta} = g^{\alpha\beta} - (P^\alpha{\bar n}^\beta +  {\bar n}^\alpha P^\beta)/{{\bar n}\cdot P}$ and $h_T^{\alpha\beta}=n_{a,T}^\alpha n_{a,T}^\beta/|n_{a,T}^2|+g_T^{\alpha\beta}/2$, with ${\bar n}\cdot P = P^0+P^z \equiv P^+$ and $n_{a,T}^\alpha = (0,\vec{n}_{a},0)$ is the transverse component of $n_{a}^\alpha$. These two tensors help to define the normal gluon NEEC $f_{g,\rm EEC}$ and the spinning gluon NEEC $d_{g,\rm EEC}$, respectively. Similarly, we can define the gluon NEECs for the proton moving in $-\hat z$ direction with momentum ${P}'$ and energy flow direction $n_b^\alpha = (1,\sin\theta_b \cos\phi_b, \sin\theta_b \sin \phi_b, \cos\theta_b)$.
The spinning gluon NEEC $d_{g,{\rm EEC}}(\theta_a)$ originates from the interference between different helicity states. 
To generate a long range correlation between $\vec{n}_a$ and $\vec{n}_b$, we need to couple two $d_{g,{\rm EEC}}(\theta)$ from both incoming protons, resulting into a $\cos(2\phi)$ asymmetry, where $\phi=\phi_a-\phi_b$. 

The spinning gluon distributions of the nucleon has also been studied in the literature under different context. In the generalized parton distribution (GPD) framework~\cite{Ji:1996ek, Muller:1994ses, Ji:1996nm, Radyushkin:1997ki}, the spinning gluon GPD, also called helicity-flip gluon GPD, predicts a $\cos(2\phi)$ asymmetry in the exclusive processes~\cite{Diehl:1997bu,Hoodbhoy:1998vm,Belitsky:2000jk}. Meanwhile, in the transverse momentum dependent (TMD) formalism, the spinning gluon distribution, referred as the linearly polarized gluon distribution, leads to a $\cos(2\phi)$ asymmetry in the associated TMD processes~\cite{Diehl:1997bu,Hoodbhoy:1998vm,Belitsky:2000jk,Boer:2010zf, Metz:2011wb, Pisano:2013cya, Hatta:2020bgy,Hatta:2021jcd,Esha:2022ovp,Caucal:2023fsf}. More recently, the $\cos(2\phi)$ asymmetry has also been discussed in the context of jet substructure~\cite{Chen:2020adz, Chen:2021gdk, Karlberg:2021kwr,Yu:2021zmw}.
The comparison of these measurements will help us understand the QCD dynamics associated with the spinning gluon.

\begin{figure}[htbp]
  \begin{center}
   \includegraphics[scale=0.65]{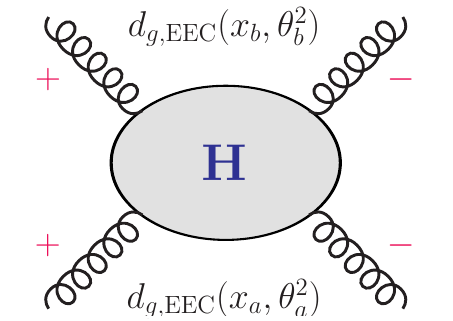} 
\caption{Long range $\cos(2\phi)$ asymmetry comes from the interference between double helicity-flip amplitudes in the partonic scattering processes. }
  \label{fg:feyn}
 \end{center}
  \vspace{-5.ex}
\end{figure}

\textbf{\textit{NEEC for Higgs Boson and top quark pair processes at the LHC.}}
The factorization for NEEC in $pp$ collisions is similar to that for the DIS processes~\cite{Li:2023gkh}. As shown in Fig.~\ref{fg:measure}, we measure the energy flows 
in $2$ arbitrary pixels on the calorimeter located at ${\vec n}_{a} =\sin\theta_a(\cos\phi_a,\sin \phi_a)$ and ${\vec n}_{b} = \sin\theta_b(\cos\phi_b,\sin\phi_b)$. The polar angles are measured with respect to the $z$-axis, i.e., the particular rapidities, and the azimuthal angles are measured from the transverse plane perpendicular to the beam direction. 
We require each of the two pixels much closer to one of the hadron beams. Therefore, these two particles are in opposite directions, forward/backward in the Lab frame, e.g., $\theta_a\to 0 $ and $\theta_b\to \pi$. The generic cross section measurement takes the following form
\bea\label{eq:e3c} 
 \Sigma(Q^2; \theta_{a,b},\phi)&=&\sum_{ij}\int  d\sigma(Q^2)\frac{E_i}{E_P} \frac{E_j}{E_P} {\cal F}(\phi; \vec{n}_{a,b})  \nn \\ 
&& ~~~~~~\times
\delta(\vec{n}_a-\vec{n}_i)\delta(\vec{n}_b-\vec{n}_j)  \,,
\eea 
where ${\cal F}(\phi; \vec{n}_{a,b})$ imposes the phase space measurement to construct $\phi$.  
In the above equation, $d\sigma(Q)$ represents partonic scattering cross section. Following previous examples, the factorization formula can be written as,
\bea\label{eq:sig-z} 
&& 
\Sigma(Q^2;\theta_{a,b},\phi)\nonumber\\
&&
= \int d\Omega \left\{x_af_{g,{\rm EEC}}\left(x_a,\theta_a^2  \right)x_bf_{g,{\rm EEC}}\left(x_b,\theta_b^2  \right) \hat{\sigma}_0\right.\\
&&\left.+x_ad_{g,{\rm EEC}}\left(x_a,\theta_a^2  \right)x_bd_{g,{\rm EEC}}\left(x_b,\theta_b^2  \right)\hat\sigma_2(Q^2)\cos(2\phi)\right\} \ ,\nonumber
\eea 
where $Q^2=x_ax_bS_{pp}$ with $S_{pp}$ the center of mass energy squared, $d\Omega$ represents additional phase space integral. $\hat\sigma_{0,2}$ are partonic cross sections calculable perturbatively. The $\cos(2\phi)$ term $\hat\sigma_2$ comes from the interference between double helicity-flip amplitudes where both incoming gluons have the same helicity as illustrated in Fig.~\ref{fg:feyn}.

Of course, the final results of $\cos(2\phi)$ asymmetries also depend on the NEEC gluon distributions. When $P^+ \theta_a \gg \Lambda_{\rm QCD}$, they can be computed from perturbative QCD with collinear splitting contributions,  
\bea 
&&
f_{g,{\rm EEC}}(x,\theta_a^2)  
= 
\frac{\alpha_s }{2\pi}
\frac{1}{\theta_a^2} 
\int_x^1 \frac{dz}{z}  
\frac{x(1-z)}{z}   \nn \\ 
&&  \hspace{10.ex}
\times
 \left[ {\cal P}_{g/q}(z)
f_q\left(\frac{x}{z} \right) 
+  {\cal P}_{g/g}(z) f_g\left(\frac{x}{z} \right)
\right]  \,, \label{eq:splitting0}\\
&&
d_{g,{\rm EEC}}(x,\theta_a^2)  
= 
\frac{\alpha_s }{2\pi}
\frac{1}{\theta_a^2} 
\int_x^1 \frac{dz}{z}  
\frac{x(1-z)}{z}   \nn \\ 
&&  \hspace{10.ex}
\times
 \frac{2(1-z)}{z}  \left[ C_F
f_q\left(\frac{x}{z} \right) 
+  C_A f_g\left(\frac{x}{z} \right)
\right]  \,, \label{eq:splitting2}
\eea 
where ${\cal P}_{g/q}$ and ${\cal P}_{g/g}$ are usual collinear splitting kernels. Additional DGLAP resummation will modify the power behavior, for which we expect a similar effect for both $f_{g,\rm EEC}$ and $d_{g,\rm EEC}$~\cite{Chen:2020adz}. 
In the following, we will apply the above leading order results to demonstrate the azimuthal asymmetries for the Higgs Boson production and top quark pair production in $pp$ collisions at the LHC.

To study the spinning gluon effect at the LHC, the simplest process is the Higgs Boson production. Similar to the TMD case calculated before~\cite{Sun:2011iw,Boer:2011kf}, the Higgs Boson can couple to the spinning gluons directly, and at the leading order 
    \begin{equation}
        \hat\sigma_2=\hat\sigma_0=\pi g_\phi^2/64\ ,
    \end{equation}
where $g_\phi$ represents the coupling between the Higgs Boson and the gluon fields in the effective theory ${\cal L}_{eff}=-\frac{1}{4}g_\phi\Phi F^{a}_{\mu\nu}F^{a\mu\nu} $~\cite{Dawson:1990zj}. The above shows that the $\cos(2\phi)$ asymmetry for Higgs production is positive and can reach a sizable value depending on the ratio between $d_{g,\rm EEC}$ and $f_{g\rm EEC}$. 
A similar $\cos(2\phi)$ asymmetry has also been found for Higgs plus two jets production, where $\phi$ is the azimuthal angle between the two jets~\cite{Plehn:2001nj}. In the common kinematics, the physics behind these two $\cos(2\phi)$ is the same, originating from the spinning gluon.

On the other hand, for the top quark pair production, $\hat\sigma_2$ is different from $\hat\sigma_0$, 
\begin{eqnarray}
         \hat \sigma_0&=&\frac{\alpha_s^2\pi}{\hat s^2}
         \left[\frac{1}{6}\frac{1}{\hat t_1\hat u_1}-\frac{3}{8}\frac{1}{\hat s^2}\right]
        \left[\hat t_1^2+\hat u_1^2+4m_t^2\hat s-\frac{4m_t^4\hat s^2}{\hat t_1\hat u_1}\right]\nonumber\\
        \hat \sigma_2&=&\frac{\alpha_s^2\pi}{\hat s^2}
        \left[\frac{3}{8}\frac{1}{\hat s^2}-\frac{1}{6}\frac{1}{\hat t_1\hat u_1}\right]
        \frac{2m_t^4\hat s^2}{\hat t_1\hat u_1}\ ,
\end{eqnarray}
for the dominant $gg\to t\bar t$ channel, where $\hat t_1=\hat t-m_t^2$ and $\hat u_1=\hat u-m_t^2$, $\hat s$, $\hat t$ and $\hat u$ are usual Mandelstam variables. Contrary to the Higgs case, the $\cos(2\phi)$ asymmetry for top quark pair production is negative. Interestingly, the asymmetry will reach the maximum value when the pair are close to the threshold where $\hat s=4m_t^2$. 
The opposite sign of $\cos(2\phi)$ asymmetries between these two processes is due to their difference in parity: Higgs Boson is parity even while the threshold top quark pair is parity odd. As a result, they couple to the incoming gluon helicity states with a different sign between $|++\rangle$ and $|--\rangle$, which leads to the opposite $\cos(2\phi)$ asymmetries. 

\begin{figure}[htbp]
  \begin{center}
   \includegraphics[scale=1]{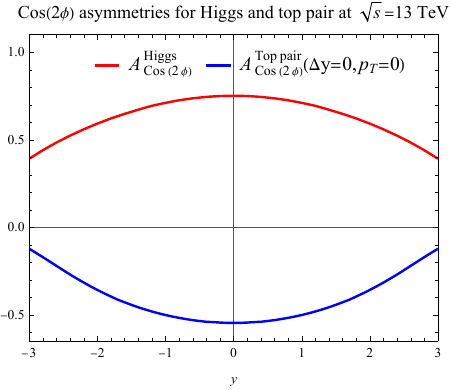} 
\caption{Long range $\cos(2\phi)$ azimuthal asymmetries associated with Higgs boson production and top quark pair threshold production as functions of their rapidity $y$. }
  \label{fg:higgstop}
 \end{center}
  \vspace{-5.ex}
\end{figure}

In Fig.~\ref{fg:higgstop}, we show the $\cos(2\phi)$ asymmetries as functions of rapidity in Higgs boson production and threshold top quark pair production. From this plot, we find that both asymmetries are quite sizable at mid-rapidity. They decrease with rapidity, which reflects $x$-dependence of the spinning gluon and the normal gluon distributions as described in Eqs.~(\ref{eq:splitting0},\ref{eq:splitting2}). The experiment measurements of these asymmetries will provide important constraints on the gluon spinning effects.

\begin{figure}[htbp]
  \begin{center}
   \includegraphics[scale=1]{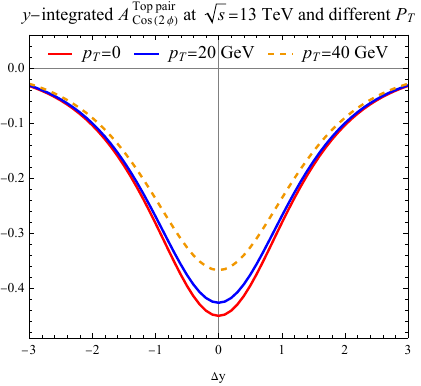} 
\caption{$\cos(2\phi)$ azimuthal asymmetries in the NEEC observable associated with top quark pair production as functions of the rapidity difference between the pair $\Delta y$ at different $p_T$. }
  \label{fg:top}
 \end{center}
   \vspace{-5.ex}
\end{figure}

As shown in Fig.~\ref{fg:top}, the $\cos(2\phi)$ asymmetry also depends on the transverse momentum and the rapidity difference between the pair $\Delta y=y_t-y_{\bar t}$ for the top quark pair. We have also computed two-photon production through the gluon-gluon fusion process by applying the amplitudes derived in the literature~\cite{Dicus:1987fk,Bern:2001df,Qiu:2011ai} and the $\cos(2\phi)$ asymmetry is smaller as compared to that of Higgs Boson production with an opposite sign.  

These results demonstrate that the $\cos(2\phi)$ asymmetries can provide a strong case to study the spinning gluon physics at the LHC. More importantly, this shall open a new avenue to study precision physics in the SM. It may also lead to a unique probe of new physics beyond the SM. Especially, the asymmetries crucially depend on the couplings between the gluon fields with different helicities and the Higgs Boson, which have been argued to be sensitive to the new physics beyond the SM and similar studies on the TMD related observable have been investigated in Refs.~\cite{Boer:2011kf,Boer:2013fca}.

Although the above results are based on the leading order calculations, we expect that higher order corrections will not modify the large $\cos(2\phi)$ asymmetries for the above processes. 
For example, in Higgs production at one-loop order, the soft gluon radiation leads to the same double logarithmic contributions for both $\sigma_2$ and $\sigma_0$. As a result, the $\cos(2\phi)$ will remain the same after resummation. 
Studies on azimuthal asymmetry between the jets in Higgs plus two jets production  
found mild dependence on both higher $\alpha_s$ order corrections~\cite{Campbell:2006xx} and parton showers~\cite{DelDuca:2006hk}. Therefore, we anticipate this attribute to persist for NEEC.

\textbf{\textit{Quantum entanglement and test of Bell inequality.}}
The $\cos(2\phi)$  correlation can be interpreted as a signature of entanglement. 
In an experiment, what is being measured are the real particles that hit the forward detectors. Although these forward-moving particles never come into contact, they remain entangled in their helicities. 
The physics picture is as follows.  
Two pairs of entangled real particles and virtual gluons are created though the splitting of the incoming partons. At this point, the two real particles are separable entities. The virtual gluons will participate the partonic hard process, while the real particles will travel towards the forward detectors at opposite ends of the beam with large momentum $E \sim P_z  \gg P_t \sim E \theta$. Once the hard process entangles the virtual gluons, it can be demonstrated that the two real particles become entangled instantaneously. 
Notably, the two real particles are always separated by space-like intervals, since $\Delta s^2 = \Delta t^2 - \Delta r_t^2 - \Delta z^2 \sim \frac{1}{E^2} - \frac{1}{E^2\theta^2} - \frac{1}{P_z^2} < 0 $, due to $\theta \ll 1$ in the forward limit. This implies the causal disconnection between the detected particles, and any non-trivial correlation between them should signify their entanglement. This remarkable property also allows us to conduct localized measurements when testing the Bell inequality. 

This observation provides a basis for testing Bell's theorem~\cite{Bell:1964kc} through the $\cos(2\phi)$ correlation. 
Leveraging the NEEC in Eq.~(\ref{eq:sig-z}), one can formulate the Bell observable 
\bea 
S(\phi_a,\phi_b) \equiv 
\frac{
\Sigma(\phi_a,\phi_b) + \Sigma(\phi_a',\phi_b')
- \Sigma(\phi_a',\phi_b) - \Sigma(\phi_a,\phi_b')}
{\Sigma(\phi_a,\phi_b) + \Sigma(\phi_a',\phi_b')
+ \Sigma(\phi_a',\phi_b) + \Sigma(\phi_a,\phi_b')} \nonumber \\ 
\label{eq:correlation}
\eea 
where $\phi_a$ and $\phi_b$ are azimuthal angles of the energy flow directed towards the detector, measured with respect to arbitrary reference vectors $r_{a,b}$. 
$\phi' = \phi+ \frac{\pi}{2}$ and can be regarded as one measures the azimuthal angles with the reference vectors perpendicular to $r_{a,b}$. 
For appropriate choices of $\phi_{a,b}$, $\tilde{\phi}_{a,b}$,
the 
Clauser-Horne-Shimony-Holt (CHSH) inequality~\cite{Clauser:1969ny}, an equivalent version of the Bell's original inequality,  
\bea\label{eq:chsh} 
B\equiv |S(\phi_a,\phi_b) - 
S(\phi_a,\tilde{\phi}_b) 
+ 
S(\tilde{\phi}_a,\phi_b) 
+ 
S(\tilde{\phi}_a,\tilde{\phi}_b) | \le 2 \,. \nonumber \\ 
\eea 
can potentially be violated. The maximum violation of the CHSH inequality for any quantum state is given by the Tsirelson's bound, $B_{\rm max} = 2\sqrt{2} \approx 2.828$~\cite{Cirelson:1980ry}.  A proof of Eq.~\eqref{eq:chsh} can be found in the supplementary material.

Fig.~\ref{fg:higgs-bell} demonstrates the concept by measuring the CHSH inequality in Eq.~(\ref{eq:chsh}) using NEEC factorization in Eq.~(\ref{eq:sig-z}).  We choose  
$\phi_a = 0$, $\phi_b = \frac{\pi}{8}$, $\tilde{\phi}_a = \frac{\pi}{4}$ and $\tilde{\phi}_b = \frac{3\pi}{8}$. Violation of the CHSH inequality is observed for the Higgs rapidity $y_{\rm Higgs} < 0.5$. We note that the significance can be dramatically improved by quark jet tagging as manifest from Fig.~\ref{fg:higgs-bell} where the CHSH inequality violation is observed for both Higgs and $t{\bar t}$ threshold production. 
We also check that increasing the machine energy leads to a more significant violation, reaching $B \approx 2.36$ for $y_{\rm Higgs} = 0$ at $\sqrt{S_{pp}}= 33\>{\rm TeV}$ without jet tagging, as the entanglement between the detected forward-moving particles intensifies near small $x$ values.

\begin{figure}[htbp]
  \begin{center}
  \includegraphics[scale=1.]{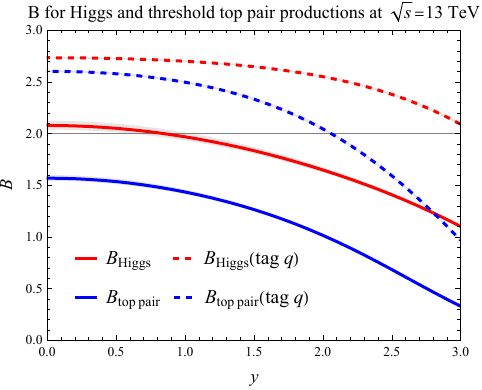}
\caption{Violation of the CHSH inequality in Higgs (red) and top pair (blue) production at the LHC. Quark jet tagging (dashed lines) substantially enhances the significance. 
}
  \label{fg:higgs-bell}
 \end{center}
  \vspace{-5.ex}
\end{figure}

\textbf{\textit{Conclusion.}} In summary, we studied the long-range azimuthal angular correlations in NEEC measurements in $pp$ collisions at the LHC. For a number of processes, we found significant large $\cos(2\phi)$ asymmetries. The comparison between these and future studies at the EIC will provide an important test of the universality of the NEEC distribution functions. Because of large asymmetries in these processes at the LHC, we emphasize that this will also open a new avenue to study precision physics for the Standard Model (SM), in particular, through comparison between Higgs Boson production, top quark pair production, and two-photon production. 

The connection between the $\cos(2\phi)$ correlation and the entanglement makes the long-range correlation in NEEC a promising alternative for testing the Bell Inequality, one of the most fundamental properties of quantum mechanics. We demonstrate the feasibility of this approach using Higgs and threshold $t{\bar t}$ production at the LHC in which violation of the Bell
inequality is significant when we perform quark jet tagging.   
Compared to the other collider-based tests discussed in the literature~\cite{Fabbrichesi:2021npl,Severi:2021cnj,Barr:2021zcp,Aguilar-Saavedra:2022uye,Aguilar-Saavedra:2022mpg,Ashby-Pickering:2022umy,Fabbrichesi:2023cev,Aguilar-Saavedra:2023hss,Bi:2023uop,Han:2023fci,Ma:2023yvd,Bernal:2024xhm,Barr:2024djo,Maltoni:2024csn}, 
the long-range correlation in NEEC enables, for the first time, a test of this fundamental quantum property in confined quantities like gluons.  
Our method benefits from the NEEC factorization theorem, ensuring the test remains local, thus closing the major potential loophole~\cite{Barr:2024djo} present in LHC-based tests. Moreover, unlike previous proposals that often require reconstructing the full kinematics which is usually challenging at the LHC, the NEEC measurement only requires determining the azimuthal angles of the energy flow deposit at the forward detectors, making it more practical for experimental implementation.

Looking ahead, extending this research to other QCD processes, including multi-jet production, and heavy quarkonium production, will be interesting to follow.  Additionally, recent investigations~\cite{Kharzeev:2017qzs,Beane:2018oxh,Tu:2019ouv,Mueller:2019qqj,Gong:2021bcp,Armesto:2019mna,Kharzeev:2021nzh,Low:2021ufv,Carena:2023vjc,Sakurai:2023nsc,Hentschinski:2023izh} have indicated that the quantum entanglement may bring novel perspectives into nuclear and particle physics. We thus anticipate our work may spark similar endeavors in unraveling the nucleon structures using the quantum information properties. These studies will promise to yield deeper insights into the effects of spinning gluons, complement our current understanding, and potentially reveal new physics beyond the SM.

\begin{acknowledgments}

\textbf{\textit{Acknowledgement.}} We thank Meng Xiao for discussions on jet tagging. 
This work is supported by the Natural Science Foundation of China under contract No.~12175016 (X.~L.), the Office of Science of the U.S. Department of Energy under Contract No. DE-AC02-05CH11231 and under the umbrella of the Quark-Gluon Tomography (QGT) Topical Collaboration with Award DE-SC0023646 (Y.G and F.Y.), the Startup Grant of Peking University, and the Asian Young Scientist Fellowship (H.~X.~Z.). 

 \end{acknowledgments}

\bibliographystyle{h-physrev}   
\bibliography{refs}

\begin{widetext}
\section*{Supplemental Material: Proof of the CHSH inequality in Eq.~(9)
}

For the convenience of the readers, we provide a self-contained proof of the CHSH inequality in Eq.~\eqref{eq:chsh}. First, we note that the two-particle NEEC defined in Eq.~\eqref{eq:correlation} satisfies
\begin{align}
    |S(\phi_a, \phi_b)| = &\ \frac{
|\Sigma(\phi_a,\phi_b) + \Sigma(\phi_a',\phi_b')
- \Sigma(\phi_a',\phi_b) - \Sigma(\phi_a,\phi_b')|}
{|\Sigma(\phi_a,\phi_b) + \Sigma(\phi_a',\phi_b')
+ \Sigma(\phi_a',\phi_b) + \Sigma(\phi_a,\phi_b')|}
\nn
\\
= &\ 
\frac{
|\Sigma(\phi_a,\phi_b) + \Sigma(\phi_a',\phi_b')
- \Sigma(\phi_a',\phi_b) - \Sigma(\phi_a,\phi_b')|}
{|\Sigma(\phi_a,\phi_b)| + |\Sigma(\phi_a',\phi_b')|
+ |\Sigma(\phi_a',\phi_b)| + |\Sigma(\phi_a,\phi_b')|}
\nn
\\
\leq &\ 
\frac{
|\Sigma(\phi_a,\phi_b)| + |\Sigma(\phi_a',\phi_b')|
 + |- \Sigma(\phi_a',\phi_b)| + | - \Sigma(\phi_a,\phi_b')|}
{|\Sigma(\phi_a,\phi_b)| + |\Sigma(\phi_a',\phi_b')|
+ |\Sigma(\phi_a',\phi_b)| + |\Sigma(\phi_a,\phi_b')|}
\nn
\\
 =&\ 1 \,.
\end{align}
Therefore, $-1 \leq S(\phi_a, \phi_b) \leq 1$. For the particular two-particle NEEC generated by the spinning gluon effects, where  
$\Sigma(\phi_a,\phi_b) = A_0 + A_2 \cos(2(\phi_a -  \phi_b))$ and $S(\phi_a,\phi_b) = \frac{A_2}{A_0}\cos(2(\phi_a-\phi_b))$ for $\phi_{a,b}' = \phi_{a,b}+\frac{\pi}{2}$, the inequality can be saturated when $A_2 = \pm A_0$. The key assumption in the proof of Eq.~\eqref{eq:chsh} is that the two-particle correlation is induced through a set of hidden variables, collectively denoted as $\lambda$,
\begin{equation}
    S(\phi_a, \phi_b)  
    = \int d\lambda\, \rho(\lambda) S_\lambda(\phi_a,\phi_b) = 
    \int d\lambda\, \rho(\lambda) R_a(\phi_a, \lambda) R_b(\phi_b, \lambda) \,,
    \label{eq:S}
\end{equation}
 where the last equation holds for local measurement, i.e., the measurement events are separated by a space-like interval and the result of a measurement on particle $a$ be unaffected by operations on the distant particle $b$, and vice versa. Here, 
\bea
R_a(\phi_a,\lambda) = \int d\phi_b S_\lambda(\phi_a,\phi_b) \,,\qquad 
R_b(\phi_a,\lambda) = \int d\phi_a S_\lambda(\phi_a,\phi_b) \,,\quad 
\eea 
where we note that $R_b(\phi_b, \lambda) \ne R_a(\phi_a,\lambda)$ in general, 
and 
\begin{equation}
    \int \rho(\lambda) d\lambda  = 1 \,, \qquad \rho(\lambda) \geq 0 \,, \qquad |R_{a,b}(\phi, \lambda)| \leq 1 \,.
\end{equation}
We proceed by first prove that
\begin{equation}
    |S(\phi_a, \phi_b) - S(\phi_a, \tilde\phi_b)| +  |S(\tilde\phi_a, \phi_b) + S(\tilde\phi_a, \tilde\phi_b)| \leq 2 \,.
    \label{eq:lemma}
\end{equation}
Using \eqref{eq:S}, we have
\begin{align}
    &\   |S(\phi_a, \phi_b) - S(\phi_a, \tilde\phi_b)| +  |S(\tilde\phi_a, \phi_b) + S(\tilde\phi_a, \tilde\phi_b)|  
    \nn
      \\
      =&\ 
      \int d\lambda\, \rho(\lambda) \Big(
      |R_a(\phi_a, \lambda)| |R_b(\phi_b, \lambda) - R_b(\tilde \phi_b, \lambda)|
      + |R_a(\tilde \phi_a, \lambda)| |R_b(\phi_b, \lambda) + R_b(\tilde \phi_b, \lambda)| \Big)
      \nn
      \\
      \leq &\ 
      \int d\lambda\, \rho(\lambda) \Big(
       |R_b(\phi_b, \lambda) - R_b(\tilde \phi_b, \lambda)|
      +  |R_b(\phi_b, \lambda) + R_b(\tilde \phi_b, \lambda)| \Big) \,.
\end{align}
Without loss of generality, suppose that $R_b(\phi_b, \lambda) \geq |R_b(\tilde\phi_b, \lambda)| \geq 0$. Then 
\begin{equation}
    |R_b(\phi_b, \lambda) - R_b(\tilde \phi_b, \lambda)|
      +  |R_b(\phi_b, \lambda) + R_b(\tilde \phi_b, \lambda)| = R_b(\phi_b, \lambda) - R_b(\tilde \phi_b, \lambda)
      +  R_b(\phi_b, \lambda) + R_b(\tilde \phi_b, \lambda) = 2 R_b(\phi_b, \lambda) \leq 2 \,.
\end{equation}
Then we have
\begin{equation}
     |S(\phi_a, \phi_b) - S(\phi_a, \tilde\phi_b)| +  |S(\tilde\phi_a, \phi_b) + S(\tilde\phi_a, \tilde\phi_b)|   \leq 2 \int \rho(\lambda) d\lambda = 2 \,,
\end{equation}
and \eqref{eq:lemma} is proved. From \eqref{eq:S}, we can apply the triangular inequality to obtain
\begin{equation}
    |S(\phi_a, \phi_b) - S(\phi_a, \tilde\phi_b) +  S(\tilde\phi_a, \phi_b) + S(\tilde\phi_a, \tilde\phi_b)|  \leq |S(\phi_a, \phi_b) - S(\phi_a, \tilde\phi_b)| +  |S(\tilde\phi_a, \phi_b) + S(\tilde\phi_a, \tilde\phi_b)|  \leq 2 \,.
\end{equation}
This completes the proof of \eqref{eq:chsh}.

\end{widetext}

\end{document}